\documentclass[sigconf]{acmart}

\usepackage{booktabs} % For formal tables
\usepackage[inline]{trackchanges}
\usepackage{balance}
\usepackage{amssymb}
\usepackage{amsthm}
\usepackage{mathtools}
\usepackage{colortbl}
\usepackage{enumitem}

\addeditor{JECA}
\addeditor{XIAO}
\addeditor{IVAN}
\addeditor{DJOLE}

\acmConference[]{}{}{} 
\begin{document}
\title[Deeply Supervised Semantic Model for CTR Prediction in Sponsored Search]{Deeply Supervised Semantic Model for Click-Through Rate Prediction in Sponsored Search}
	
	\author{Jelena Gligorijevic}\authornote{Co-first author}
	\affiliation{%
		\institution{Temple University}
		\streetaddress{1925 N 12th St}
		\city{Philadelphia} 
		\state{Pennsylvania} 
		\country{US}
		\postcode{19122}
	}
	\email{jelena.stojanovic@temple.edu}
	
	\author{Djordje Gligorijevic}\authornotemark[1]
	\affiliation{
		\institution{Temple University}
		\streetaddress{1925 N 12th St}
		\city{Philadelphia} 
		\state{Pennsylvania} 
		\country{US}
		\postcode{19122}
	}
	\email{gligorijevic@temple.edu}
	
	\author{Ivan Stojkovic}
	\affiliation{
		\institution{Temple University}
		\streetaddress{1925 N 12th St}
		\city{Philadelphia} 
		\state{Pennsylvania} 
		\country{US}
		\postcode{19122}
	}
	\email{ivan.stojkovic@temple.edu}
	
	\author{Xiao Bai}
	\affiliation{
		\institution{Yahoo! Research}
		\streetaddress{701 1st Avenue}
		\city{Sunnyvale} 
		\state{California}
		\country{US}
		\postcode{94089}
	}
	\email{xbai@oath.com}
	
	\author{Amit Goyal}
	\affiliation{
		\institution{Criteo}
		\streetaddress{325 Lytton Avenue}
		\city{Palo Alto} 
		\state{California}
		\country{US}
		\postcode{94301}
	}
	\email{a.goyal@criteo.com}
	
	\author{Zoran Obradovic} 
	\affiliation{
		\institution{Temple University}
		\streetaddress{1925 N 12th St}
		\city{Philadelphia} 
		\state{Pennsylvania} 
		\country{US}
		\postcode{19122}
	}
	\email{zoran.obradovic@temple.edu}

	% The default list of authors is too long for headers}
	\renewcommand{\shortauthors}{J. Gligorijevic et al.}

	\begin{abstract}
%	\change{Sponsored search is a major online monetization model for commercial web search engines. Ads that are relevant to a query are usually ranked based on the expected revenue if they are clicked by users. Therefore, it} 
 
% It is critical for sponsored search engines to match the ads that are relevant to a query, and to accurately predict the likelihood of them being clicked for the query. 
In sponsored search it is critical to match ads that are relevant to a query and to accurately predict their likelihood of being clicked.
Commercial search engines typically use machine learning models for both query-ad relevance matching and click-through-rate (CTR) prediction. 
%However, most matching models only rely on syntactic and semantic features to determine the similarity between a query and an ad, ignoring the fact that highly relevant ads may not attract any click in practice. On the other hand, widely adopted click models heavily rely on click history to make predictions, which limits their generalizability for new queries and ads that have little click history. %In this work [IVAN]
However, matching models are based on the similarity between a query and an ad, ignoring the fact that a retrieved ad may not attract clicks, while click models rely on click history, being of limited use for new queries and ads.
We propose a deeply supervised architecture that jointly learns the semantic embeddings of a query and an ad as well as their corresponding CTR.
		%\cBl The proposed architecture takes the texts of a query and an ad as inputs to learn (1) an embedding for each of them using bi-directional recurrent neural networks and attention networks, and (2) the CTR using convolutional neutral networks.\cB Two loss functions that are specific to semantic matching and CTR prediction are combined for joint optimization. 
        We also propose a novel cohort negative sampling technique for learning implicit negative signals. 
        We trained the proposed architecture using one billion query-ad pairs from a major commercial web search engine. This architecture improves the best-performing baseline deep neural architectures by 2\% of AUC for CTR prediction and by statistically significant 0.5\% of NDCG for %statistically significant 
        query-ad matching. 
		%The proposed cohort negative sampling reduces the training time by \% compared to the alternative technique. 
	\end{abstract}
	
	%
	% The code below should be generated by the tool at
	% http://dl.acm.org/ccs.cfm
	% Please copy and paste the code instead of the example below. 
	%
	
	\begin{CCSXML}
		<ccs2012>
		<concept>
		<concept_id>10002951.10003260.10003272.10003273</concept_id>
		<concept_desc>Information systems~Sponsored search advertising</concept_desc>
		<concept_significance>500</concept_significance>
		</concept>
		<concept>
		<concept_id>10002951.10003317.10003359.10003361</concept_id>
		<concept_desc>Information systems~Relevance assessment</concept_desc>
		<concept_significance>300</concept_significance>
		</concept>
		<concept>
		<concept_id>10010147.10010257.10010293.10010294</concept_id>
		<concept_desc>Computing methodologies~Neural networks</concept_desc>
		<concept_significance>500</concept_significance>
		</concept>
		</ccs2012>
	\end{CCSXML}
	\ccsdesc[500]{Information systems~Sponsored search advertising}
	%\ccsdesc[300]{Information systems~Relevance assessment} [IVAN]
	\ccsdesc[300]{Computing methodologies~Neural networks}

	\keywords{Deep Learning, CTR Prediction, Query to Ad Matching}
	 %, Sponsored Search[IVAN]
	
	\maketitle
	
	\section{Introduction}
	
	Sponsored search has been a major monetization model for commercial web search engines, contributing a significant portion to the multi-billion dollar industry of online advertising. Given a query, it is critical for search engines to retrieve relevant ads and to accurately predict their CTR in order to maximize the expected revenue while ensuring good user experience. Both overpredicting and underpredicting CTR would result in revenue loss.% to search engines as they would either allocate limited search result slots to unattractive ads or miss opportunities of showing attractive ads.
    
% In sponsored search, ads that are relevant to user queries are usually shown on the north, south and east of organic search results. These ads are monetized on a pay-per-click model: ads are ranked based on their expected revenue, computed as the predicted click-through-rate (CTR). Advertisers only pay for clicked ads based on a generalized second price auction \cite{king2007internet}. Therefore, given a query, it is very critical for search engines to retrieve relevant ads and to accurately predict their CTR in order to maximize the expected revenue while ensuring good user experience. Both overpredicting and underpredicting CTR would result in revenue loss to search engines as they would either allocate limited search result slots to unattractive ads or miss opportunities of showing attractive ads.
	
	Machine learning models made great success in predicting CTR for sponsored search. Most of the models adopted in the industry rely on a large set of well-designed features to predict CTR. Features extracted from click history have been proved very effective \cite{cheng2010personalized}. However, models that heavily rely on click features often fail to generalize to new queries and new ads with insufficient history \cite{richardson2007predicting}. To make predictions in such cases, models resort to syntactic or semantic features extracted from queries, ads, and advertisers~\cite{richardson2007predicting,li2014semantic}. Deep neural networks were also proposed to learn features from traditional models \cite{jiang2016research} or to learn CTR from existing features \cite{zhang2014sequential}. In spite of the existing success, designing and selecting appropriate features remains a very challenging problem for CTR prediction~\cite{he2014practical}. %that have little history [IVAN] shaparenko2009data,
	
	Following the progress of deep learning in natural language processing, recent efforts rely on deep neural networks to capture semantic similarities between queries and ads to predict CTR without any feature engineering \cite{edizel2017deep}. Such models are learned end-to-end from clicks without explicit supervision for capturing the semantic similarity between a query and an ad, and as we show in this work, they have not achieved their full potential in CTR prediction. %However, since [IVAN]
	
	A number of recent works \cite{grbovic2016sigir,jaech2017match} used deep neural networks to model the semantic similarity between a query and an ad. These models were shown effective in a query to ad relevance matching. However, as they do not directly model clicks, retrieved ads are only weakly correlated to the ads presented to users based on expected revenue (which highly depends on the predicted CTR). 
	
	In this work, we propose a deeply supervised end-to-end architecture for CTR prediction in sponsored search. This architecture jointly learns CTR and discriminative representations of queries and ads such that clicked query-ad pairs are also mapped closer in the embedded space. Specifically, this architecture takes the texts of a query and an ad as input to bi-directional recurrent neural networks (bi-RNNs) and attention networks to learn discriminative distributed embeddings. Query and ad embeddings are then matched together and fed into convolutional neural networks (CNNs) to predict CTR. Two losses, specific to semantic matching and CTR prediction, are jointly optimized at different levels of the architecture to provide a deep supervision for both tasks. This architecture has the advantages of \textit{(i)} not relying on any feature engineering; \textit{(ii)} directly optimizing CTR prediction; \textit{(iii)} directly learning semantic representations to enable query-ad matchings more correlated with clicks and expected revenue. The key contributions are: %of this work are as follows. [IVAN]
	
	\begin{itemize}[leftmargin=0.4cm]
		\item We propose a novel deep architecture that jointly learns CTR and discriminative representations of queries and ads. To the best of our knowledge, this is the first attempt to simultaneously learn CTR and semantic embeddings using click data. By optimizing two logistic losses specific to CTR prediction and semantic matching instead of using only one CTR specific logistic loss, we were able to achieve statistically significant lift in AUC. 
		\item We propose a novel cohort negative sampling technique that naturally draws information from implicit negative signals in the data. We assess the impact of this technique in terms of performance and prove the convergence of our method through theoretical analysis.  %\change[DJORDJE]{that significantly reduces the efforts of drawing negative samples during training}{...}
		\item We conduct an extensive empirical evaluation of the proposed architecture using about one billion query-ad samples from the Yahoo! web search engine. Comparison with state-of-the-art CTR prediction models shows that our model improves the AUC of the best-performing baseline model by 2\%.  
		\item We evaluate the quality of the query and ad embeddings learned by our model through a query-ad matching task using a large-scale editorially labeled dataset. Comparison with state-of-the-art matching models shows that our model improves the  NDCG  of the best-performing baseline by statistically significant  0.5\%, confirming its ability to learn meaningful semantic embedding. 
	\end{itemize}

	\section{Related Work}
	\label{sec:related_work}
	%In this section, [IVAN]
    We first present problems and challenges in sponsored search and review most recent advances in deep learning approaches. Subsequently, we review other relevant advances in deep learning, which have previously been applied only on tasks different than ours. 
	
	\subsection{Related Work in Sponsored Search}
	The frequently tackled problems of improving the sponsored search include CTR prediction, query rewriting and query to ad matching.
    
	A large body of work focused on predicting probability that an ad would be clicked, if shown as a response to a submitted query \cite{graepel2010web,mcmahan2013ad,he2014practical}. State-of-the-art approaches have mainly used handcrafted features of ad impressions obtained from historical impressions (i.e. ad and query CTR's, users' historical features, etc.) and semantic similarities of queries and ads \cite{richardson2007predicting}. These approaches range from Bayesian \cite{graepel2010web} to feature selection approaches \cite{he2014practical}, however, a common challenge for all is creating and maintaining a large number of sparse contextual and semantic features \cite{mcmahan2013ad}. % manually handcrafted sounds like oxymoron [IVAN]
	
	Focusing on the broad matching of queries and ads that have similar semantic meaning is another line of research \cite{fuxman2008using}. The task is to retrieve ads that are semantically similar to the query \cite{grbovic2016sigir} without exactly matching keywords (i.e. query ``running machine'' and ad ``elliptical trainer''). This task has been commonly addressed by query rewriting models \cite{jones2006generating} or by semantic matching \cite{grbovic2016sigir,fuxman2008using,huang2013learning}. %another interesting line of research [IVAN]
	
	More recently, many approaches for CTR prediction utilize various deep learning techniques. Deep learning primarily alleviates issues of creating and maintaining handcrafted features by learning them automatically from the ``raw'' query and ad text data.
	
    % DJORDJE: I removed this as it refers to sequence modeling we are not dealing with
% 	Web sessions (users' clicks and views on the Web during a predefined period of time) pose an information rich source of users' behavior patterns that can be used to predict clicks. RNNs were used to predict clicks based on the sequence of users' activities~\cite{zhang2014sequential}. Using such sequences has been successfully employed for learning semantic representations of queries and ads with neural embeddings, which was then used for query to ad matching.
	It is common to learn query and ad semantics from ad impressions for a given query with click information. In~\cite{huang2013learning} authors proposed a deep structured semantic model (DSSM) with dual architecture that embedded a query on the one side and an ad on the other and learned matching between the two given the click information. In order to improve quality of the learned semantic match and capturing query intent, a word attention mechanism was successfully used for the query and ad representations \cite{zhai2016deepintent}.
	
	Some of the approaches are defined as a CTR prediction task rather than as a matching task. In \cite{shan2016deep}, features of an impression (query text, ad text, ad landing page, campaign ID, keywords, etc.) are learned automatically from the impression, in a deep architecture, to predict click probability. Other models,  DeepMatch~\cite{edizel2017deep} and MatchTensor~\cite{jaech2017match} proposed very deep dual network architectures for query and ad embeddings with a matching layer to learn ad impression representations useful for CTR prediction. % Several models, namely DM... [IVAN] 
	
	Both groups of approaches, learning semantics of queries and ads and learning to predict CTR are widely used in systems for serving ads. However, they pose a trade-off, while semantic learning learns relations between queries and ads, it has no direct click probability notion, CTR prediction models, on the other hand, may suffer from not capturing the semantics of queries and ads implicitly thus affecting their prediction quality. The approach we propose in this study is a well-rounded framework for ad systems capable of both learning quality semantics of queries and ads as well as being able to accurately predict click probability. 
	
	The two mentioned approaches, DeepMatch and MatchTensor have shown great results in practice and will, thus, be the main baselines and building blocks for the model proposed in this study. The two approaches are conceptually very similar as both learn independent representations of a query and an ad, and use a matching layer to associate their words, and finally learn to predict CTR. However, the difference between them is in way they learn representations of words, i.e. DeepMatch primarily uses temporal convolutional layers, while MatchTensor uses bi-RNNs. Also, they propose slightly different matching layers, DeepMatch proposes a cross-feature matrix, while MatchTensor proposes cross-feature tensor. As both models perform exceptionally well, we present a detailed analysis of performance of both models experimentally in Section~\ref{sec:experiments}. 
	
	The model proposed in this study further extends on the advances described above by addressing their shortcomings by introducing novel ways of learning semantically rich representations. As such, the proposed model demonstrates the state-of-the-art results on both CTR prediction and query2ad matching tasks, traditionally modeled by different families of models. This is achieved by means of (i) learning new blocks in the deep architectures to improve modeling capacity, (ii) adding deep supervision to improve quality of learned representations deep in the model and (iii) learning parameters in an efficient and information-rich way to capture more of the available semantics in the dataset.
	\vspace{-5pt}
	\subsection{Related Work in Deep Learning}
	\label{sec:related}
%	Lastly, it is important to discuss advances in deep architectures capable of learning very discriminative and abstract representations of texts relevant for the proposed approach
	
	Many approaches for mathematical characterization of language, that model sequence data, were proposed to advance the field of natural language processing. Initially, distributed low-dimensional representations of words were introduced in \cite{rumelhart1988learning} and recently successfully applied for learning semantic and syntactic relations among words \cite{mikolov2013efficient}. The idea of using distributed representations of words was further exploited in approaches as RNNs, capable of learning an embedded high-dimensional representation of sequences.% or tokens [IVAN]
	
	\textbf{\textit{Recurrent Neural Networks. }}
	RNNs are a popular family of models for sequential problems. While previous approaches have often modeled word sequence as an order-oblivious sum, RNNs learn representations of word sequences by maintaining internal states, which are updated sequentially and are used as a proxy for predicting the target. The ability to stack multiple layers allows building deeper representations that result in great improvements on many tasks. In particular, an architecture of RNNs called long short-term memory (LSTM) cell achieved the biggest success~\cite{greff2017lstm}. 
% 	More recently, the sequence-to-sequence paradigm for RNNs was proposed, where the input sequence is encoded using the ``encoder'' network, and the output sequence is generated using the ``decoder'' network \cite{sutskever2014sequence}. This paradigm has been successfully used for translating sentences from one language into another. 
	
	\textbf{\textit{Attention Network Models. }}
%	Building on the recent success of RNN's \cite{sutskever2014sequence}, 
    Attention models dynamically re-weight the importance of various elements (words, phrases or characters) in the text during the decoding process, thus altering the learned representation. Use of attention demonstrated considerable improvements in performance~\cite{bahdanau2014neural}. An attention mechanism was developed as a separate neural network that takes a sequence of word embeddings and learns attention scores for each word, where more ``important'' words in the document have higher attention leading to a more focused higher-order representation of the sequence. Attention models were recently adapted for the general setting of learning compact representations of documents \cite{zhai2016deepintent}. %(i.e. focusing attention) 
	
	\textbf{\textit{bi-RNNs. }}	
    Another successful paradigm is the bi-RNN, where two RNNs (i.e. LSTM, thus bi-LSTM) independently encode the text sequence in both forward and in backward direction~\cite{schuster1997} computing representation that captures complex relations between words in the text. Final sentence representation is obtained by aggregating representations of the two single-directional LSTMs, and it was observed that bi-LSTM's perform well on datasets where there is no strict order in the sequences, such as the case with Web queries.

	\textbf{\textit{Convolutional Text Models. }}
	Recently, architectures for sequence modeling increasingly include temporal convolutions as building blocks. Temporal convolutions are capable of learning representations of sequences which proved as a good building block for several deep architectures. Good examples being ConvNet for text classification \cite{zhang2015character} and the Very Deep CNN (VDCNN) model \cite{conneau2016very}, both of which use temporal convolutions to model a sequence of words/characters with aim to perform classification. These models successfully outperformed RNN based models. In this study, we use word-level VDCNN as one of the baselines, as it consists of equivalent blocks as the DeepMatch model, save the matching layer.

\textbf{\textit{Deeply supervised models. }} 
Recently, several models drew benefits from utilizing deep supervision \cite{zhang2016augmenting,lee2015deeply,szegedy2015going}. The key idea is to use supervision at various layers across the model to enforce discriminativeness of the features \cite{lee2015deeply} and potentially resolve exploding/vanishing gradients \cite{zhang2016augmenting,szegedy2015going}. However, existing approaches mostly use the same predictive task in deeper layers as in the final layer \cite{lee2015deeply,szegedy2015going} and in some cases use reconstruction loss \cite{zhang2014sequential}. We build upon these advances proposing a novel approach of using deep supervision specifically designed to extract information from the data in an explicit way, which would not be possible otherwise.

\textbf{\textit{Learning from implicit negative signals. }}
This has for a long time been a challenging task for domains with implicit negative signals. Recently, search2vec model for learning with implicit negative signals from sponsored search sessions was proposed \cite{grbovic2015sigir} with improved performance and speed of the algorithm. Furthermore, \cite{chen2017sampling} have confirmed this approach and applied it on the special case of bipartite graphs. We exploit implicit negatives in our model and consider comparing to search2vec algorithm in Section~\ref{sec:query2ad}.

	\section{Proposed Model}
\begin{figure}[]
	\includegraphics[width=0.49\textwidth]{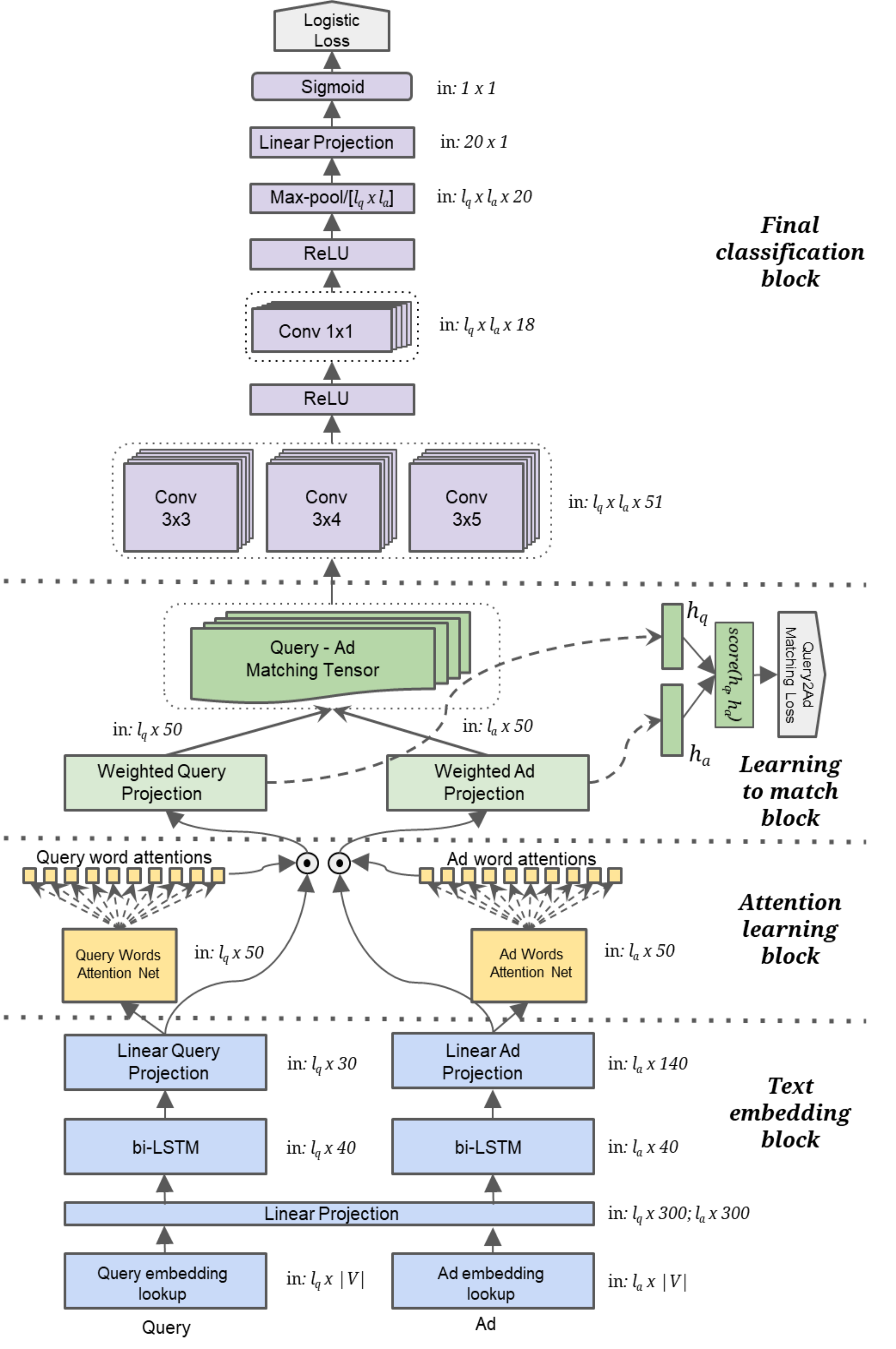}
		\caption{Proposed DSM model block diagram}
		\label{fig:dstm}
        \vspace{-10pt}
	\end{figure}

	Graphical representation of the proposed model, which we call the Deeply Supervised Matching (DSM) model is given in Figure~\ref{fig:dstm}.
    %leverages supervised information of click data to learn better representations of queries and ads jointly by modeling CTR prediction loss
    
	% The main choice of architecture was drawn from the MatchTensor model~\cite{jaech2017match}, due to its excellent performance in our experiments. 
	The model takes query text and ad text as inputs, and it learns their separate embeddings through a series of layers, including bi-direction LSTM and attention layers. Learned embeddings are then used in two-fold matching: (1) embeddings of query and ad words are used in an elementwise product to construct a matching tensor, and  (2) matching of dense representations of query and ad is learned using a novel matching loss designed for sponsored search. Learned matching tensor is then passed through series of convolutional and pooling blocks to learn CTR prediction.%\cBl This complex and deep architecture with multiple losses enables our model to perform outstandingly on several tasks.\cB [IVAN]
	
	%A detailed description of the modeling blocks as well as the similarities and distinctions to existing approaches is given below.

	\subsection{Blocks of the proposed model}

%	Our proposed model consist of the following parts: [IVAN]
    %blocks/parts/functionalities:

	\subsubsection{Query and Ad text embedding}
	Embeddings of query and ad texts are done in two networks. First $l_q$ words in the query and $l_a$ words in the ads are embedded into a $d_{qa}^{(1)} = 300$ dimensional space. Then, a fully connected layer is used to learn linear combinations of words in a $d_{qa}^{(2)} = 40$ dimensional space. These two layers share weights for both queries and ads. % 40 space [IVAN] performed -> done
    %\change[XIAO]{$l_a$ ads}{
	% In several published approaches, authors use small dimensionality for the word embedding \cite{edizel2017deep}, while in our initial experiments we observed increase in prediction quality when words were given higher dimensional representation. To prevent overblowing of parameters we limited the dimensionality of the first word embeddings to $300$.
	Embeddings of query and ad are passed to the respective bi-LSTM layers such that the model learns complex relations between words, which is in particular important for queries that may have a different order of words but the same meaning (i.e. ``best restaurants in Boston'' vs. ``Boston best restaurants''). Due to different lengths of query and ad text embedding sizes are now $d_q^{(3)} = 30$ and $d_a^{(3)} = 140$, as suggested in the literature \cite{jaech2017match,zhai2016deepintent}. Finally, fully connected layers are used to reduce representations of all words in the same, reduced, dimensional space $d_{qa}^{(4)} = 50$, resulting in representations $v_q = l_q \times d_{qa}^{(4)}$ and $v_a = l_a \times d_{qa}^{(4)}$, for query and ad, respectively.
	% (rather than single directional relations)
	\subsubsection{Attention learning}
	In order to learn rich representations of queries and ads, it is imperative to focus on words that carry the most information. In order to learn representations that focus on important parts of queries and ads we employ the attention models from machine translation and adapt them to a more general case of using word scores for learning compact (vector) representations \cite{zhai2016deepintent}.
	Two attention blocks are used, one for query text and one for ad text. These blocks yield word scores, that signify attentions the model will give to different words. Both attention models are implemented as two-layered individual neural networks $s_q(v_q; \theta_q)$ and $s_a(v_a; \theta_a)$ with softmax at their final layer
	{\small\begin{equation}
		t_q^{(i)} = \frac{exp(s_q(v_q^{(i)}; \theta_q))}{\sum_{i = 1}^{l_n} exp(s_q(v_q^{(i)}; \theta_q))}.
		\end{equation}}
	Neural networks $s_q(v_q^{(i)}; \theta_q)$ and $s_a(v_a^{(i)}; \theta_a)$ learn real valued scores for each $i^{th}$ word in a given query and ad, respectively. Attentions learning in DSM is coupled with the entire network (end-to-end). % , allowing for an end... [IVAN]
	
	Attentions $t_q^{(i)}$ for a query word, and $t_a^{(i)}$ for an ad word, are then used to re-weight their input representations $v_q$ and $v_a$ to obtain compact representations of query and ad used for learning to match as $h_q = \sum_{i}t_q^{(i)}*v_q^{(i)}$ and $h_a = \sum_{i}t_a^{(i)}*v_a^{(i)}$.
	There are other ways of obtaining compact representations $h_q$ and $h_a$, such as sum, average or max of individual word vectors. However, our experiments, as well as available literature \cite{zhai2016deepintent}, demonstrate that such strategies are inferior to using attention.
	
	\subsubsection{Query and Ad matching}
	%As discussed before [IVAN]
    Many models for sponsored search advertising have either the capability to learn good quality semantic representations of queries and ads, or the capability to perform CTR prediction well without explicitly modeling semantics, thus (over-)specializing in only one of the tasks. %, \textcolor{blue}{often yielding in an overfit}. 
    %Xiao: Does ctr prediction without explicit modeling semantics yield to overfit? The above is not very clear.
	To address this, we have two matching processes in our framework.%: an implicit and an explicit one.
	
	First, similarly to MatchTensor~\cite{jaech2017match}, we build a tensor for implicitly matching words in a query and an ad. $l_q$ words in a query and $l_a$ words in an ad, with $d_{qa}^{(4)}$-dimensional embeddings, are matched in a cross product tensor of shape $l_q \times l_a \times d_{qa}^{(4)}$. Each word in a query will be matched to each word in an ad, and the element-wise product of their vectors will be a thread in the matching tensor. Finally, an exact-match $l_q \times l_a$ slice is added to the tensor, with all zeros except for words that co-occur in a query and an ad, where we put ones. This slice serves as a bias and yields slight improvement as opposed to the model that does not use exact matches \cite{jaech2017match}.
	
    % Further more
	Second, we propose explicit matching to capture semantic similarity between a query and an ad. We propose a way to match the vectors $h_q$ and $h_a$, where we aim to embed them such that they are closer in the embedded space if there was a click and farther away if there was no click, similarly to \cite{grbovic2016sigir}. % wish -> aim [IVAN]
	To achieve this, we optimize scores between $h_q$ and $h_a$ vectors, where scores are posed as an inner product of the vectors. %implemented -> posed [IVAN]
	To avoid introducing the computational complexity of negative sampling, we introduce a cohort negative sampling approach to optimize the matching function. The detailed description, as well as convergence analysis of the proposed optimization strategy, are given in Sections~\ref{sec:batch_negative_sampling} and ~\ref{sec:sgd}.
	
	Benefits of using multiple learning tasks for the same model have recently been recognized~\cite{lee2015deeply}. Deep models benefit from enforcing the middle layers to be discriminative, which is beneficial for the final predictive task, as discriminative classifiers trained on highly discriminative features will perform better than a discriminative classifier trained on less discriminative features. In our case, representations of query and ad should be close for semantically similar pairs and distant for dissimilar ones. Such representations benefit the classification task as the semantic relations have been well captured deep in the model. Due to adding such deep supervision, our model is named the Deeply Supervised Matching (DSM) model. % discovered -> recognized [IVAN]
	% Authors show that introducing multiple loss functions for the middle layers can benefit the overall quality of predictions for the entire model.
	% For now, it should be noted that we add one more task for model to learn deep in the architecture to ensure that learned query and ad representations are semantically discriminative, which would not be achieved previously. 

	\subsubsection{Learning to predict from matched representation}
	The matching tensor from the previous block is then convolved through the entire depth $d_{qa}^{(4)} + 1$ by three convolutional blocks with different filter sizes: 3 for query words; and 3, 4, and 5 words for ad filters. The number of filters is fixed to 6 for the first set of convolution blocks and 20 for the final convolutional layer. % filters [IVAN]
	Complex representations between a query and ad words are learned here, and they are passed through the ReLU layer, after which another $1 \time 1$ convolution with ReLU was used before the two-dimensional max-pool layer that embeds the whole query-ad impression in a single vector. Finally, the vector is fed to a fully connected layer and passed through a sigmoid layer $\sigma(\cdot)$ to obtain the logits of the model.%(rectified linear unit)   [IVAN] 

	\subsection{Logistic and Matching Losses}
	Finally, to optimize the parameters of DSM, we have logistic loss $\mathcal{P}$ for the CTR prediction based on logits from the topmost layer:% obtained [IVAN]
	\begin{equation}
	\mathcal{P}(W) = -\frac{1}{N} \sum_{n=1}^{N}(y_n log(\hat{y}_n) + (1-y_n)log(1-\hat{y}_n) ),
	\label{eq:logistic_loss}
	\end{equation}
	where $\hat{y}_n$ are obtained logits after final sigmoid layers and $y_n$ is click label for the $n^{th}$ ad impression. 
	The matching loss $\mathcal{Q}$ for query and ad vectors, as a negative sampling approximation, can be generalized as a composition of positive and negative pairs \cite{mikolov2013distributed}: % straightforwardly generalized [IVAN]
% \change[JECA]{PROMENA FORMULE}{pogledaj staru i proveri ostale da se usklade}
% \begin{equation}
% 	\label{eq:batch_negative_sampling}
% 	\begin{split}
% 	\mathcal{Q}(W) &= \sum_{b=1}^{B}(\sum_{j \in \mathcal{D}_p} \mathcal{Q}^{+}(W) + \sum_{k \in \mathcal{D}_n} \mathcal{Q}^{-}(W) \large) \\ &= \sum_{b=1}^{B} \large( \sum_{j \in \mathcal{D}_p} -\log\sigma(h_q^{(b)T} h_a^{(j)}) + \sum_{k \in \mathcal{D}_n} \log\sigma(-h_q^{(b)T} h_a^{(k)}) \large),
% 	\end{split}
% 	\end{equation}

\begin{equation}
\label{eq:batch_negative_sampling}
\begin{split}
\mathcal{Q}(W) &= \sum_{b=1}^{B}(\sum_{j \in \mathcal{D}_p^{(b)}} \mathcal{Q}^{+}(W) + \sum_{k \in \mathcal{D}_n^{(b)}} \mathcal{Q}^{-}(W) \large) \\ &= \sum_{b=1}^{B} \large( \sum_{j \in \mathcal{D}_p^{(b)}} -\log\sigma(h_q^{(j)T} h_a^{(j)}) + \sum_{k \in \mathcal{D}_n^{(b)}} \log\sigma(-h_q^{(k)T} h_a^{(k)}) \large),
	\end{split}
	\end{equation}
	where $B$ is the total number of batches, while $\mathcal{D}_p$ and $\mathcal{D}_n$ are positive and negative impressions within each batch, respectively. In our implementation, we use a variant of the negative sampling loss for learning to match query and ad vectors, called cohort\footnote{We use word cohort to disambiguate our sampling strategy from the traditional mini-batch i.i.d. sampling.} negative sampling. 
	As will be discussed later in the paper, this loss differs from the negative sampling loss proposed in \cite{mikolov2013distributed}, as negative samples are used within the cohort but not sampled ad-hoc, thus saving computational time.
	% Equation~\ref{eq:batch_negative_sampling} can be further generalized as a composition of positive and negative pairs:
	% \begin{equation}
	% \mathcal{Q}(W) = \sum_{b=1}^{B}(\sum_{j \in \mathcal{D}_p} \mathcal{Q}^{+}(W) + \sum_{k \in \mathcal{D}_n} \mathcal{Q}^{-}(W) \large).
	% \end{equation}
	
	\noindent
	The final loss function becomes the sum of Eq.~\ref{eq:logistic_loss} and Eq.~\ref{eq:batch_negative_sampling}
	\begin{equation}
	\mathcal{L}(W) = \mathcal{P}(W) + \mathcal{Q}(W).
	\label{eq:loss_function}
	\end{equation}
	We use $W$ to annotate the set of all parameters in the DSM. %\cR Formulation~\ref{eq:loss_function} is similar to multi task loss \cite{liu2016deep} but for fundamentally different task?! \cB % reasons -> task [IVAN]
	
	Based on the Lemma 1 in~\cite{lee2015deeply}, a good solution for $\mathcal{Q}$ is also a good solution for $\mathcal{P}$. However, conversely is not necessarily true. This clearly states that features learned for $\mathcal{P}$ may not be optimal for $\mathcal{Q}$. In the case of our application, features learned for the classification task may not capture semantic similarities between queries and ads that may carry considerable amounts of information.
	Another interesting aspect of using multiple optimization functions is that it is reasonable to assume that $\mathcal{L}$ and $\mathcal{P}$ share the same optimum~\cite{lee2015deeply}, while $\mathcal{Q}$ can be observed as a regularizer. 
	% It was shown that surrogate losses, such as the $\mathcal{Q}$ in our case, drop their gradient much faster than the final loss. To validate this, we have experimented with different strategies of normalizing the two looses and their gradients using batch normalization \cite{ioffe2015batch}. Results of the analysis are briefly discussed in Section~\ref{sec:normalizing_losses_experiment}.
	
	Therefore, it is important to notice that $\mathcal{Q}$ is not used for learning to match explicitly, but as stated before, to enforce discriminative embeddings of the lower layers such that final logits reflect semantic information found in the data. To demonstrate this, we used the DSM model for query to ad matching and compared it to well-established models for the task in Section~\ref{sec:query2ad}.
	
	Weights are initialized by a truncated normal initializer. To optimize $\mathcal{L}$, we use Adam~\cite{kingma2014adam} with a decaying gradient step.% the objective L [IVAN] optimizer
	
	% \change[DJOLE]{Batch Negative Sampling}{Cohort negative sampling!}
	\subsubsection{Cohort Negative Sampling for Matching Loss}
	\label{sec:batch_negative_sampling}
	
\begin{figure}[ht!]
		\includegraphics[width=0.3\textwidth]{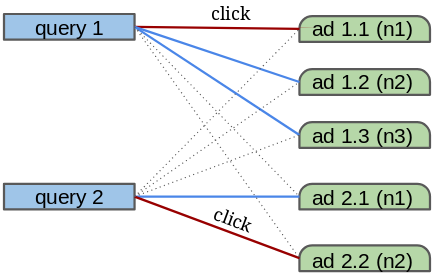}
		\caption{Cohort negative sampling (an example with queries and served ads in the position ``north'', n1 up to n5) Red links are ad clicks, blue links are ads displayed but not clicked, and negative pairs we create by coupling queries and ads that were not displayed for that ad - dotted links.} %gray 
		\label{fig:batch_negative_sampling}
      \end{figure}

	The nature of ad serving in sponsored advertising is that for each query, the publisher (search engine in this case) can provide a set of ads on different positions on the search result page. The most impactful position is called ``north'' (ads placed above organic links) and it yields the largest click-through rate for ads \cite{chen2012position}. Up to five ads can be presented at this location (n1 -- n5), and users may or may not click on any of them. Click/No-click information provides an implicit information on a query and ad relevance that we can learn from. Thus, to learn matching we need to focus on a group of query--ad pairs that were served to the user for a given search, and we can pull several such searches in the cohort we use for training. Such data allows us to learn a semantic match of a query and an ad implicitly, based on users' feedback. In the past, learning such implicit relations between queries and ads has shown great benefit in sponsored search ad recommendations \cite{grbovic2016sigir}, while its computational benefits were supported in \cite{chen2017sampling}. In this study, unlike in \cite{chen2017sampling}, implicit negative samples naturally occur as signals from the users, furthermore they do not consider that complete ground-truth bipartite graph is needed to obtain the good working model, as artificial negative samples can be harmful if a pair is semantically related. The later issue is leveraged with matching tensor layer, while matching loss merely plays a role of discriminativeness enforcing regularizer.
	An example of a cohort of users' search query impressions used for training our models is given in Figure~\ref{fig:batch_negative_sampling}.
	
	% For each query, up to five ads can be served on different positions (we consider north positions only: n1 - n5), some of which may be clicked and other not, thus having an explicit negative link provided by the user. We thus define our mini batches as complete query ad impressions as shown in Figure~\ref{fig:batch_negative_sampling}. For a given query, all ads that were served are used to create query ad pairs. In terms of computational savings, we have no need to sample $k$ negative ads from some distribution $P_n$, thus computation is potentially decreased to $c*b$, where c is the number of queries in the batch ($c \leq b$).
	
	Traditionally, techniques such as negative sampling \cite{mikolov2013distributed} were proposed as a speedup for costly partition functions while learning to match. However, in negative sampling for each positive sample (in our case query-ad $(q,a)$ pair with click) $m$ there needs to be $k$ sampled ads from some distribution $P_n$ that provide negative pairs for a given query, thus ending up with a total of $m + m*k$ embedding operations prior to matching. In our case, we do not sample $k$ negative ads, thus the computation is decreased by $m*k$ in addition to capturing implicit signals from users.

	% In our model, we have a dual network of embedding words in queries and ads which are learned to be matched in the matching tensor, a layer that simply learns cross-word relations of query and ad, without learning how query and ad are in fact related. In the past learning relations between queries and ads has shown great benefit in sponsored search ad recommendations \cite{grbovic2016sigir}.
	% For this purpose the negative sampling \cite{mikolov2013efficient} procedure was proposed as a speedup for costly partition functions while learning to match. However, in negative sampling for each positive sample (in our case query-ad $(q,a)$ pair) $b$ there needs to be sampled $k$ ads from some distribution $P_n$ that provide negative pairs for a given query, thus ending up with total of $b(1+k)$ embedding operations prior to matching.
	
	In cohorts with insufficient negative pairs for the partition function to provide satisfactory approximation of true distribution, we resort to negative cross-referencing queries with ads that are found in cohort and were not served for those queries (dotted gray links in the Figure~\ref{fig:batch_negative_sampling}), obtaining up to $< m*(m-1)$ negative pairs.
	
	For further analysis, it is useful to characterize the matching loss function in terms of expected values over query $q$, and ad $a$ pairs as positive (click) and negative (no click) examples drawn from their respective distributions $P_d$ and $P_n$:% for further analysis, of the proposed optimization  [IVAN]
    
	\begin{equation}
	\begin{split}
	\mathcal{Q}(W) &= \mathbb{E}_{(q,a) \sim P_p(q,a)}[\mathcal{Q}^{+}(q,a; W)] + \mathbb{E}_{(q,\grave{a}) \sim P_n(q,\grave{a})}[\mathcal{Q}^{-}(q,\grave{a}; W)] \\ &= \mathbb{E}_{q \sim P_p(q)} \left[ \mathbb{E}_{a \sim P_p(a|q)}\mathcal{Q}^{+}(q,a; W)] + \mathbb{E}_{\grave{a} \sim P_n(\grave{a})}[\mathcal{Q}^{-}(q,\grave{a}; W)] \right],
	\end{split}
	\label{eq:pairwise_matching}
	\end{equation}

Due to the nature of the data, we assume that the query is observed first and then ads are provided (conditioned on the query). Obviously, we do not assume independence between two random variables. However, for sampling negative samples     $(q,\grave{a})$, we assume that the distribution $P_n(q,\grave{a}) = P_p(q)P_p(\grave{a})$ is approximately scaled by $\frac{P_n(\grave{a})}{P_p(\grave{a})}$ to allow the factorization $P_n(q,\grave{a}) = P_p(q)P_n(\grave{a})$. %, thus preserving unbiasedness \add[DJOLE]{cite or is it clear already}. 
	
	It is important to formalize the sampling strategy $\mathbb{E}_B[\mathcal{Q}_B(W^t)]$ that is different from the simple i.i.d. sampling:

\begin{equation}
	\begin{split}
	& \mathbb{E}_B[\mathcal{Q}_B(W^t)] = 
	\\ 
	& \mathbb{E}_{(q,a) \sim P_p(q,a)}[\mathcal{Q}^{+}(q,a; W)] + \mathbb{E}_{(q,a) \sim P_p(q)P_p(\grave{a})}[ \frac{P_n(\grave{a})}{P_p(\grave{a})} \mathcal{Q}^{-}(q,\grave{a}; W)] \\ 
	&+  \frac{1}{m(m-1)} \sum_{j=1}^{m(m-1)} \mathbb{E}_{(q,a) \sim P_p(q,\grave{a})}[ \frac{P_n(\grave{a})}{P_p(\grave{a})} \mathcal{Q}^{-}(q,\grave{a}; W)] = 
	\\ 
	& \mathbb{E}_{q \sim P_p(q)} [ \mathbb{E}_{a \sim P_p(a|q)}[ \mathcal{Q}^{+}(q,a; W)] + \mathbb{E}_{\grave{a} \sim P_n(\grave{a})}[\mathcal{Q}^{-}(q,\grave{a}; W)] \\ & + \mathbb{E}_{\grave{a} \sim P_p(\grave{a})}[ \frac{P_n(\grave{a})}{P_p(\grave{a})} \mathcal{Q}^{-}(q,\grave{a}; W)] ] = 
	\\ 
	& \mathbb{E}_{q \sim P_p(q)} \left[ \mathbb{E}_{a \sim P_p(a|q)}[ \mathcal{Q}^{+}(q,a; W)] + 2 \mathbb{E}_{\grave{a} \sim P_n(\grave{a})}[\mathcal{Q}^{-}(q,\grave{a}; W)] \right] = \\ & \mathcal{Q}(W^t).
	\end{split}
	\end{equation}

      %Cohort negative sampling. This is an example of a cohort with queries and served ads in the position ``north'' (n1 up to n5). Red links annotate ad clicks, while blue links annotate ads displayed but not clicked. Finally, we create negative pairs by coupling queries and ads that were not displayed for that ad - gray dotted links.
    
    We can define samples in the cohort as taking positive and negative samples from the $P_p(q,a)$ and $P_n(q, \grave{a})$ distributions, respectively. As shown in Figure~\ref{fig:batch_negative_sampling}, the first expectation is for positive, clicked pairs, the second is for ads not clicked, and the third is for negative query--ad pairs created within the cohort. Due to properties of expectation and joint distribution, we are allowed to factorize the expectation to obtain the result equivalent to Eq.~\ref{eq:pairwise_matching}. Using the gradient property of expectation, we obtain the following lemma:%Finally, 
	\begin{lemma}[Cohort negative sampling is an unbiased stochastic gradient estimator]
		Given samples in $B$ generated by the cohort negative sampling algorithm, the stochastic gradient is unbiased as the expected cohort gradient equals the true gradient: $\mathbb{E}_B[\bigtriangledown \mathcal{L}_B(W^t)] \equiv \bigtriangledown \mathcal{L}(W^t)$
		\label{lem:unbiased_estimatod}
	\end{lemma}

	\subsubsection{Stochastic Gradient Descent View}
	\label{sec:sgd}
	To optimize our objective we use the stochastic gradient descent (SGD) algorithm. The update at the $t^{th}$ iteration of SGD is in the form:
	\begin{equation}
	W^{t+1} = W^{t} - \eta_t \bigtriangledown \mathcal{L}_t(W).
	\end{equation}
	For examples chosen randomly in iteration $t$, SGD provides an unbiased estimate of the gradient: $\mathbb{E}_B[\bigtriangledown \mathcal{L}_B(W^t)] \equiv \bigtriangledown \mathcal{L}(W^t)$. 
	%For randomly chosen examples in index $t$, SGD provides an unbiased estimate of the gradient at each iteration: $\mathbb{E}_B[\bigtriangledown \mathcal{L}_B(W^t)] \equiv \bigtriangledown \mathcal{L}(W^t)$.
    
	Analyzing the convergence of the SGD algorithm for non-convex problems has been a big research question. However, it was shown in~\cite{ghadimi2013stochastic,reddi2016stochastic} that SGD follows a local convergence bound. Furthermore, \cite{ghadimi2013stochastic} provide the following theorem showing that SGD will converge within $T$ steps, based on the assumptions that $\bigtriangledown \mathcal{L}(W^t)$ is an unbiased estimator and that the expected variance of the gradient $\bigtriangledown \mathcal{L}(W^t)$ is upper-bounded by $\sigma^2$.
	\begin{theorem}
		[Local convergence of non-convex SGD~\cite{ghadimi2013stochastic,reddi2016stochastic}] Suppose $\mathcal{L}$ has a $\sigma$-bounded gradient; let $\eta_t=\eta=c/\sqrt[]{T}$ where $c=\sqrt[]{\frac{2(\mathcal{L}(W^0) - \mathcal{L}(W^*))}{L \sigma^2}}$, and $W^*$ is an optimal solution to (\ref{eq:loss_function}). Then, the iterates of SGD satisfy
		\begin{equation}
		\min_{0<t<T-1} \mathbb{E}[\parallel \bigtriangledown \mathcal{L}(W^t) \parallel^2] \leq \sqrt[]{\frac{2(\mathcal{L}(W^0) - \mathcal{L}(W^*))L}{T}} \sigma.
		\end{equation}
		\label{thrm:local_convergence}
	\end{theorem}
    \vspace{-5pt}
	\noindent
	Although $\mathcal{L}$ is a composite of $\mathcal{P}$ and $\mathcal{Q}$, the theorem still applies according to Lemma 2 from~\cite{lee2015deeply}. Thus, we only need to show that the proposed sampling strategy in Section~\ref{sec:batch_negative_sampling} yields an unbiased minimizer $\mathcal{Q}$, which follows from the Lemma~\ref{lem:unbiased_estimatod}.
	Satisfied assumptions conclude the convergence of our algorithm.
    %As the assumptions are satisfied, this concludes the analysis of convergence of our algorithm.

	\section{Experiments}
	\label{sec:experiments}
	%In order to assess the capabilities of the proposed DSM approach, 
    We conducted an extensive empirical evaluation for the CTR prediction task on a large dataset (about one billion query-ad samples) from a major commercial search engine (Section~\ref{sec:CTR}). We also evaluate the quality of the query and ad embeddings learned by our model through a query-ad matching task using a large-scale editorial labeled dataset (Section~\ref{sec:query2ad}). The data and the experimental set-up used for both tasks are described in each of these sections. %to test and  [IVAN]

	% The data will be described in the remainder of this section, as well as the analysis on the CTR prediction and Query2Ad matching tasks.
	
	% \cR
	%     We conduct extensive empirical evaluation of the proposed architecture using $\sim1$ billion query-ad samples from a major commercial search engine. 
	
	%     Comparison with state-of-the-art CTR prediction models shows that our model improves the AUC of the best-performing baseline model by \cR XXX\%\cB. We also show that learning embeddings and using two losses help to improve the AUC of our model by \cR XXX\% \cB and \cR XXX\% \cB respectively.  
	
	%     We also evaluate the quality of the query and ad embeddings learnt by our model through a query-ad matching task using a large-scale editorial labeled dataset. Comparison with state-of-the-art CTR prediction models shows that our model improves the \cR NDCG@10 \cB of the best-performing baseline model by \cR XXX\% \cB, confirming its ability of learning meaningful semantic embedding. 
	% \cB
	
	% \subsection{Experimental setup}
	% \label{sec:exp_setup}
	
    \subsection{CTR prediction}
	\label{sec:CTR}
	For the CTR prediction task, the aim is to estimate, as accurately as possible, the probability $P(click | ad, query)$ that a user would click on an ad displayed after submitting a query.
    
      \vspace{-10pt}

	\subsubsection{Click-through rate data}
	\label{sec:data}
	To train and test the proposed model and baselines for this task, we collected a random sample of logged query-ad pairs served by a popular commercial search engine. 
	The sample comprises of 987,734,146 query-ad pairs for training and 16,881,864 for testing, containing only advertisements placed at the top (north) of search result page (ads that are served above organic search links). The data consists of a query text on one side, and ad title, ad description and ad display URL on the other side. The query and ad texts are processed and normalized using an in-house tool to remove special characters and punctuations, make letters lower case, fix common typos, split URLs, etc. All example pairs are accompanied with information whether the ad was clicked or not, which we use as supervised information to train all models. 
	To better characterize the dataset, we comment on its distribution of the queries.
	% in Table~\ref{tab:freq}.
	% presents the distribution of queries in the sampled dataset. 
	% It can be seen that
    %Fix this next sentence, too confusing. What does the 90% mean in reference to the 75%?
	A majority (75\%) of queries are infrequent (tail queries), i.e. appearing less than five times overall, and if measured in the test set only there are more than 90\% them. 
	As discussed before, this is a major limitation of most of the traditional CTR prediction models, and given the volume of the tail queries, this reaffirms the necessity for predictive models that can generalize when insufficient or no click history is available. For a subset of queries that are seen often (appear more than 20 times, called head queries) we expect all the models to perform better, even though they make up only about $3\%$ of the training set and less than $1\%$ of the testing dataset.

	\subsubsection{Baselines} 
	\label{sec:baselines}
	We compare our proposed Deeply Supervised Matching (DSM) approach against several alternatives described in Section~\ref{sec:related}: A linear logistic regression learned on top of the word embedding layer (LM), Very Deep CNN (VDCNN) \cite{conneau2016very}, DeepMatch (DM) \cite{edizel2017deep}, and MatchTensor (MT) \cite{jaech2017match}. 
	
	All deep learning models were trained in two ways: (i) with the use of pre-trained word embedding vectors (obtained from~\cite{pennington2014glove}); and (ii) when the word embeddings are learned specifically for the task, directly from the training dataset. All the models were implemented in the Tensorflow framework, and were run on a distributed cluster with multiple GPU machines (Nvidia p80) due to the size of the data. 
	The initial learning rate of $0.0001$ was set for the Adam Optimizer, while the mini-batch size was set to $512$.
	%All models were trained on $4$ epochs with initial learning rate of $0.0001$. The batch size used was $512$. All models were implemented in Tensorflow on Spark and trained in batch mode using an Adam optimizer over 4 epochs.
	
	\subsubsection{Metrics} For assessing the quality of the estimated CTR probabilities, we use a common classification performance measure of area under the ROC curve (AUC), as well as  Accuracy obtained after choosing the appropriate classification threshold.  In addition, we study the bias of the predicted probabilities. Unbiasedness is a desirable property, as positive bias leads to overly-optimistic estimates and a waste of resources, and negative bias leads to overly-conservative estimates and a waste of opportunity.

	\subsubsection{Results}
\begin{figure}[h]
		\includegraphics[width=0.40\textwidth]{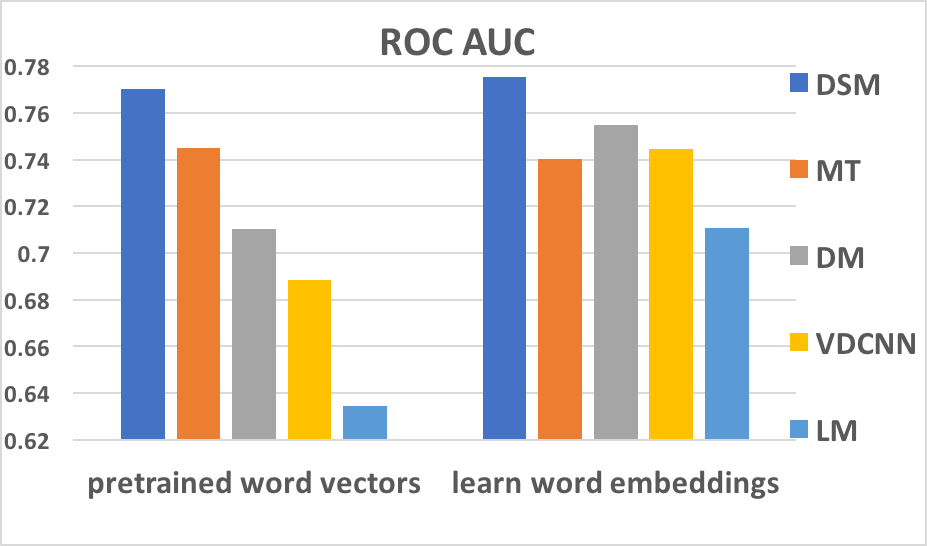}
		\caption{Models with learned embeddings (on the right) perform better than models with pretrained vectors (on the left)}\label{fig:CTRall}
	\end{figure}

	Prediction performance results, on the holdout testing dataset, are presented in Fig.~\ref{fig:CTRall}. 
	Results in the Table\ref{learned_bias} are the best results obtained by the respective model. The DSM approach outperforms all the alternatives with the highest AUC of $0.775$.
	% (all DSM model AUC's are above the line of 0.76, while all alternative models are bellow that line). 
	% Moreover, all four variations of DSM model have similar performance (low variance), suggesting robustness of the proposed deep neural architecture.
		\begin{table}[h]
		\centering
		\caption{Performance of the proposed models vs baselines}
		\label{learned_bias}
		\begin{tabular}{l|cccccc}
			\textbf{Model}    & \textbf{DSM} & \textbf{MT} & \textbf{DM} & \textbf{VDCNN} & \textbf{LM} \\
			\hline
			\hline
			\rowcolor[HTML]{EFEFEF} 
			\textbf{AUC}      & \textbf{0.775 }                  & 0.745       & 0.755       & 0.744          & 0.711       \\
			\textbf{Bias}     & \textbf{0.991}                & 1.046       & 1.033       & 0.974          & 0.965       \\
			\rowcolor[HTML]{EFEFEF} 
			\textbf{Accuracy} & \textbf{0.742 }                 & 0.703       & 0.719       & 0.734          & 0.711      
		\end{tabular}
	\end{table}

	We first evaluate the simplest way (LM) of learning to predict CTR from combined text data of query and ad, and we observe a decent performance of such an approach, which resonates well with the word embedding approaches described in Section~\ref{sec:related_work}. Furthermore, we see that by introducing deep models, such as VDCNN, we are able to achieve significant lifts in performance (3\% lift in AUC). 
	However, by introducing individual embeddings of query and ad to capture specificities of both, and learning to match the two, such as in the case of the DM or MT models, we see that the results are further improved (1\% lift in AUC). 
	Finally, when a model is capable of capturing discriminative features deep in the architecture, we obtain further improvements (additional 2\% of AUC lift).
	Accuracy measure consistently sets DSM as the best performing model.%gives the same ordering of models, empirically backing previously made assumptions about the impact of recent modeling choices.
	
	Furthermore, we evaluate the bias of predictions made by different models, and observe that the DSM model is the most unbiased model in the experiment (closest to the ratio of $1$). This implies that the expected number of clicks deviates the least from the exact number of clicked ads, thus achieving better monetization.
	The results show that the $DSM$ model's click expectation would on average be wrong for $9$ clicks, out of $1000$, which is $17$ clicks better compared to the next best $VDCNN$ model, with $26$ out of $1000$. This significantly impacts revenue due to a volume of served ads.
	
	\textbf{\textit{Learn word embeddings vs. use pre-trained word vectors. }} 
	As all baselines suggest using pretrained word embeddings in their original approaches, we examined the effect of learning embeddings in an end-to-end manner, rather than using pretrained ones. Results in Figure~\ref{fig:CTRall} show that the models where the word embeddings are learned directly on the task of CTR prediction, in a majority of cases are superior to their counterparts which use pre-trained vectors. Thus, we argue that it is important for such models to capture word specificities of the domain rather than using external embedding.% that may be suboptimal.
    
The following two experiments show results obtained by the best version (using pretrained word vectors vs. learning word embedding) of the respective model.

	% Therefore, in the remainder we will focus the analysis to models with learned embeddings.
	% \paragraph{AUC, Bias and Accuracy}
	% \cR ovo sve da se obrise \cB
	% In Table~\ref{learned_bias}, more performance indicators are given for the models with learned embeddings as better performing alternatives. 
	% From there it can be seen that the DSM model has AUC of $0.775$ and Accuracy of $0.742$, and is clearly few percents better than the alternatives.
	% A similar conclusion can be stated for the models' bias, where the DSM model has smaller bias than alternatives (it is closer to the ratio of $1$), and the less the expected number of clicks deviates from the exact number of clicked ads, the better the monetization will be.
	% %The last column in the Table~\ref{learned_bias} shows how much the achieved bias would affect the expected number of clicks (per 1 million impressions).
	% The results show that the $DSM$ model's click expectation would on average be wrong for $9$ clicks, out of $1000$, which is $17$ clicks better compared to the next best $VDCNN$ model, with $26$ out of $1000$. 
	% %\add[IVAN]{[too strong claim...]}, which means we expected to earn more money than we actually did
	
	%Table~\ref{pretrained_bias} presents same analysis for models trained using existing embedding vectors. Results are supporting conclusion about the AUC, but no longer holds for the bias of the models (we'll probably exclude this table...)

	\textbf{\textit{CTR prediction for Head, Torso and Tail Queries. }}
	It is expected that predictability of CTR depends on the query frequency. For example, for less frequent queries there may not be enough data to generalize properly. Therefore, in this subsection, we analyze the influence of the query frequency on the model predictive performance. 
	For that purpose, examples were divided into three categories: the most frequent ``head'' (">20" occurrences), least frequent ``tail'' ("<5" occurrences), and ``torso'' in-between.
    
	\begin{figure}[h]
		\includegraphics[width=0.40\textwidth]{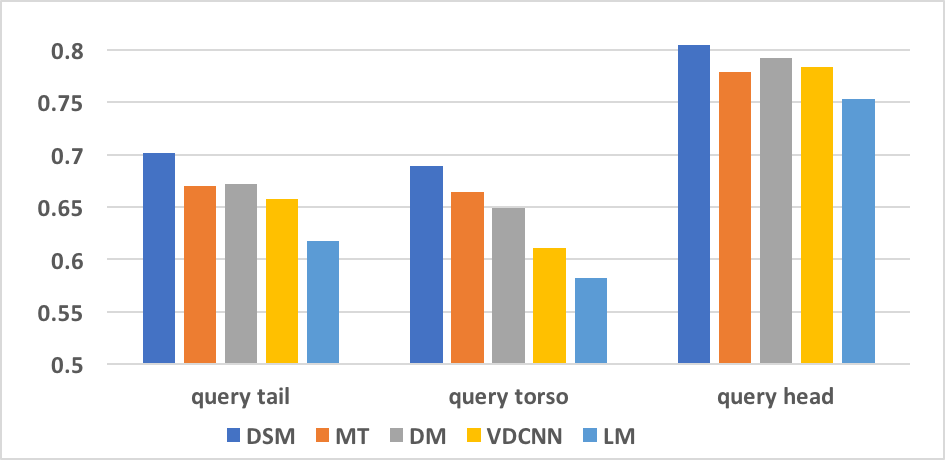}
		\caption{AUC for CTR decomposed by query frequency}
		\label{freq}
	\end{figure}
    
	Results presented in Figure~\ref{freq} align with the common sense expectation that the most frequent queries (``query head'') will be more predictable. The less frequent ``torso'' and ``tail'' queries have expectedly lower AUC (more than 10\% less than ``head''), where the least frequent queries (from ``tail'') seems to have slightly higher predictive performance, compared to the ``torso''.

\begin{figure}[]
		\includegraphics[width=0.40\textwidth]{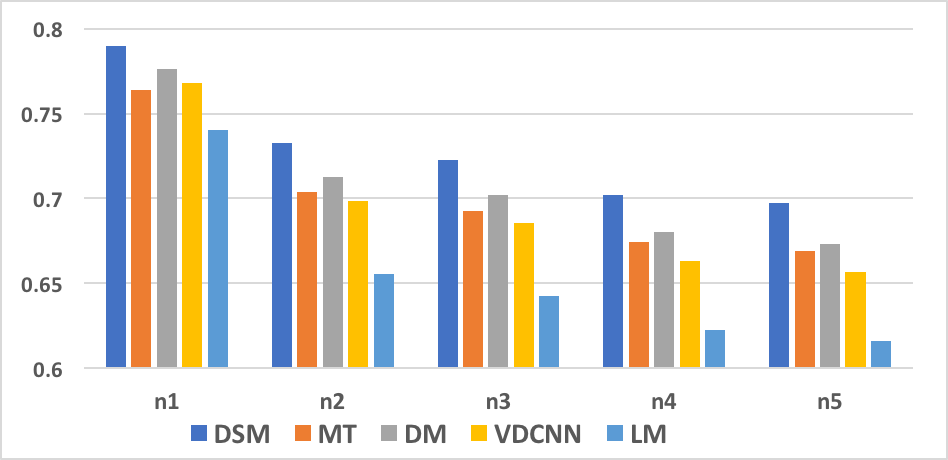}
		\caption{AUC for CTR decomposed by impression position}
		\label{posit}
	\end{figure}

	\textbf{\textit{CTR prediction over different Ad Positions. }}
	It was acknowledged that Ad position plays an important role in CTR prediction~\cite{chen2012position}. %previously observed->acknowledged [IVAN]
    For example, ads placed in the north section are more likely to be clicked than those in the south or east sections, both because it was considered the most relevant (by algorithm), and because its position is the most favorable (convenient) one. Therefore, we also analyze the influence of the ad position on the model predictive performance. %in this subsection, 
	For that purpose, we segregated the examples into 5 groups based on their positions in the north section (top one is no. 1, and up to 5, as it goes down).
	Results presented in Fig.~\ref{posit} convey that predictability decays with the rise in the position number. From the first to the second position it displays the sharpest decrease in the AUC, and from-then-on it goes more gradually until the last, fifth position. Still, the proposed model is the best on all sections.%However, the proposed model is still the best performing one on each of the sections.
    
        \begin{figure*}[]
	\includegraphics[width=0.80\textwidth]{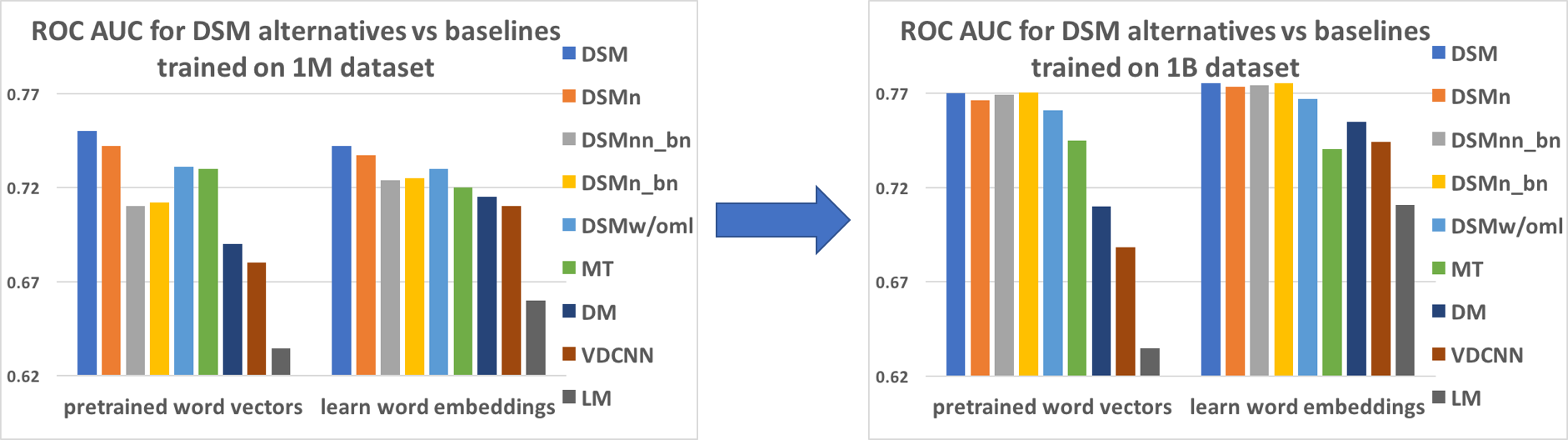}
%             \vspace{-5pt}
		\caption{Effect of Data set scale on models' CTR prediction performance}
		\label{fig:training_size}
        \vspace{-5pt}
	\end{figure*}
	\textbf{\textit{CTR prediction - training set scale impact. }}
	We also studied models training on datasets of different scales, small with millions of examples, and large with billion of examples. As shown in Fig.~\ref{fig:training_size}, scale matters when trying to characterize models for ad impression data. For example, models that use pretrained word vectors perform better on smaller dataset than their learn-embeddings alternatives, as the models that learn embeddings require more data to learn meaningful representations of words. We also note that batch normalization algorithms on smaller datasets perform much worse than their non-batch-normalizing alternatives, which is not the case on the larger dataset, suggesting that algorithms that are using batch normalization need more data to learn good representations.

    % This results clearly indicates a need for characterization and evaluation of such models, as ones considered in this study on very large data as the potential implication derived from experiments on data order of magnitude lower may be faulty.
    
    \textbf{\textit{Robustness of the DSM model. }}
	\label{sec:normalizing_losses_experiment}
	Additionally, for the proposed methodology ablation analysis, we study the effect of batch normalization, loss normalization, and attention pooling when we remove the deep supervision from the DSM model. Hence, we had four varieties of our model: plain $DSM$, $DSM$ with normalization of the two losses (trying to prevent one of the losses dropping too fast) $DSM_{n}$, DSM with batch normalization on the fully connected layers $DSM_{bn}$ to prevent large fluctuations of the logistic loss and DSM with both batch normalization and normalized losses $DSM_{n\_bn}$.

	%For our DSM approach we additionally tested how ``losses normalization'' and batch normalization~\cite{ioffe2015batch} affected the predictive performance, 
	
	Results of all exploited normalization strategies yield comparable prediction performance with $0.7754$, $0.7734$, $0.7743$ and $0.7727$ AUC for $DSM$, $DSM_n$, $DSM_{bn}$ and $DSM_{n\_bn}$, respectively.

	\textbf{\textit{Logistic loss vs. matching loss. }}
	Finally, we removed the matching loss from the model to evaluate the gain obtained by it. Furthermore, we noticed that the matching loss drops much faster than the logistic loss, even after losses normalization. That confirms that the surrogate loss served as a form of regularization~\cite{lee2015deeply} that forces the hidden layer of a query and ad representations to be semantically discriminative thus yielding higher quality CTR predictions and enabling the model to excel on matching tasks. %perform well -> excel [IVAN]
	We see a larger drop when removing the matching loss with $0.7671$ AUC (the Wilcoxon signed-rank test p--value $8.63\text{e}^{-05}$ ), thus validating that the matching loss benefits the quality of the CTR prediction.
	%statistical significance [IVAN]
	\subsection{Query2Ad Matching}
	\label{sec:query2ad}
	Finally, we assess the quality of the learned representations. The proposed DSM learn semantic matching of a query-ad pair as an effect of the matching layer and deep supervision. To validate this, we evaluate our model on the query to ad (query2ad) matching task, traditionally used for performance assessment. Note that this is not the primary task of the DSM, however, due to the nature of the proposed matching, it has the ability to perform it well. The scores between query and ad used for matching are the final layer's logits, that reflect query-ad semantics as well as the click probability.%the quality of matching models
	% To the best of our knowledge it is the first model of its kind capable to perform at the highest level for both CTR prediction and query2ad matching tasks of the computational advertising.
	
	\subsubsection{Relevance data} 
	To evaluate the quality of query and ad embeddings, we used an in-house dataset consisting of a query-ad pair that was graded editorially. The editors were instructed to grade $65,446$ query-ad pairs as either Perfectly Relevant, Highly Relevant, Relevant, Somewhat Relevant, Barely Relevant, or Irrelevant as in \cite{aiello2016role}. For each ad, the editors had access to ad title, description, and display URL to help them reach their judgment. For each query ($8,315$ unique queries) there was on average $\sim7$ graded ads, allowing us to evaluate ranking of ads in addition to relevance.

	\subsubsection{Baselines}
	We compared our method to traditional relevance models: Gradient boosted decision trees (with 1000 trees) \cite{zheng2008general} ($GBDT_{1000}$), with 185 text-based features \cite{aiello2016role} (trained on $700,000$ editorial query-ad pairs) and the $BM_{25}$ \cite{robertson1994some}. We also use other CTR prediction task baselines (described in Section~\ref{sec:baselines}), where, as for the DSM, logits of the models were used as matching scores. %These CTR prediction models are compared to show the ability of DSM to learn semantic matching between query and ad simultaneously as learning CTR.
    %probabilistic relevance method for retrieval
	Finally, we evaluated the search2vec \cite{grbovic2016sigir} for the matching task. Since the model is only trained for known queries and clicked ads, the coverage of the model on our editorial dataset was small (2,167 unique queries coming from 8,725 query ad pairs out of 65,446 records) and as such model yielded only fairish results ( [0.7, 0.8] for NDCG@2 to NDCG@7), so we do not show them in Figure~\ref{fig:NDCG_K5_65K}. %However, [IVAN]
    
    For matching quality, we use $precision@K$ and Normalized Discounted Cumulative Gain $NDCG@K$ \cite{wang2013theoretical} averaged across all queries.
	%To evaluate matching quality of models, we use $precision@K$ and Normalized Discounted Cumulative Gain $NDCG@K$ \cite{wang2013theoretical} averaged across all queries.
	
%	 \subsubsection{Metrics}
%	 To evaluate matching quality of models, we use $precision@K$ and Normalized Discounted Cumulative Gain $NDCG@K$. The $precision@K$ is computed for each query as a fraction of relevant (clicked) ads within the $K$ retrieved ones, and average precision is reported. The $NDCG@K$ \cite{wang2013theoretical} further measures how well the ranked scores align with the ranked editorial grades, and averaged results across all queries are reported.
	
	\subsubsection{Matching Results}
	
      \begin{figure}[H]
		\includegraphics[width=0.38\textwidth]{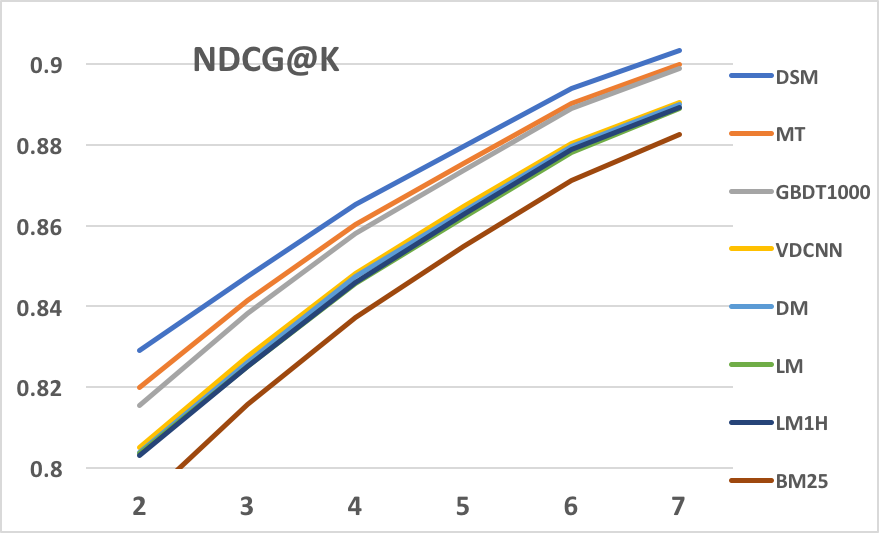}
		\caption{NDCG@K on editorial 65K query-ad pairs}\label{fig:NDCG_K5_65K}
	\end{figure} % measured judgments of 
    
	\textbf{\textit{NDCG. }}  Relevance was assessed using the $NDCG@K$ \cite{wang2013theoretical}, and the results are given in Figure~\ref{fig:NDCG_K5_65K}. We observe that the DSM approach improves over the alternatives (higher values of NDCG). Even though the difference is not obvious because of the $NDCG@2$ to $NDCG@7$ scores' scale, Wilcoxon signed-rank test p-value of $2.69e^{-05}$ measured on $NDCG@1$ to $NDCG@100$ shows that the improvement of the DSM model over alternatives is statistically significant. DSM improves the NDCG@7 of the GBDT model by 2\% and the best deep learning baseline MT by 0.5\%.
    In addition, we measure Precision@K for all the models, but for the lack of space we report here statistically significant average improvement of 1.5\% over the next best alternative.  
    
\begin{figure}[!h]
 		\includegraphics[width=0.40\textwidth]{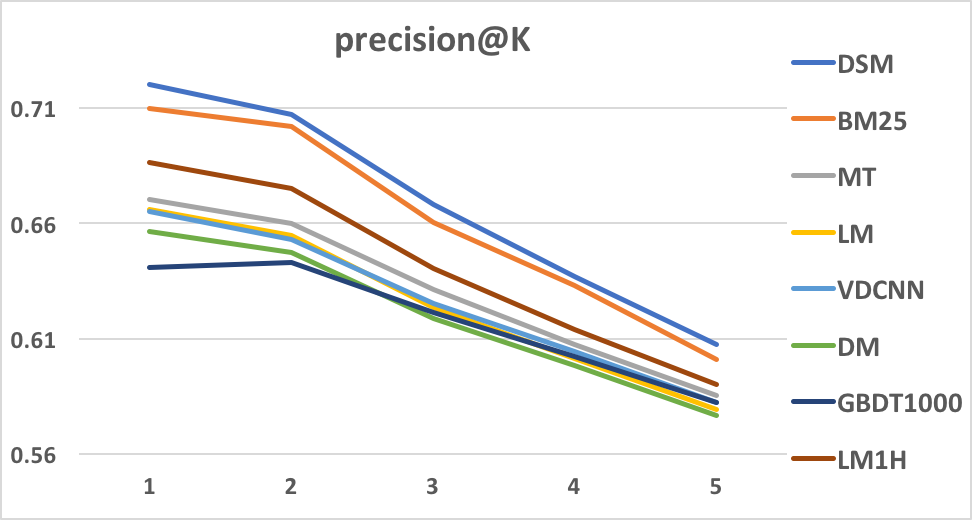}
 		\caption{precision@K measured on editorial judgments
 			of 65K query-ad pairs}\label{fig:precision_K5_65K}
 	\end{figure}

 	\textbf{\textit{Precision. }}  We also measure Precision@K to further characterize models, as shown in Figure~\ref{fig:precision_K5_65K}. The $DSM$ model is still the best performing model. However for this metric, traditional $BM_{25}$ model performs as the second best model. Statistical significance test of the improvement of the $DSM$ over the $BM_{25}$ model returns p-value of $8.85e^{-05}$, confirming that observed improvements are indeed statistically significant. %Similar traits of results are observed for the Recall@K, where the $DSM$ was best performing model, followed by the $BM_{25}$ model, with successful statistical significance test, with p-value of $XXX$.

\section{Final Remarks}
The results of our extensive experiments demonstrate that the proposed DSM model outperforms state-of-the-art approaches on CTR prediction tasks, as measured by multiple metrics. It was the most accurate, and had the least bias of all the approaches. Our model also outperformed other competitive algorithms on a query to ad matching task, as measured by the NDCG. Ablation study confirmed that the dual loss architecture (statistically significant) enhances the model performance. Moreover, our DSM model was  the best performer over different scales of data, frequencies of the queries, ad positions and embedding choices.
Above mentioned suggests that joint training of two complementary tasks, as query to ad matching and CTR prediction are, through deep supervision, yields high quality, versatile models.    

%Above mentioned suggests that the two tasks of sponsored search are complementary, and that modeling them jointly yield good performing versatile models.
    %We proposed a novel model for sponsored search advertising. Our theoretical and empirical analysis has shown that it is the best performing model yet for learning to predict CTR using only available text data. Furthermore, we have shown that it also performs well on query2ad matching task, traditionally tackled by a different family of models. Finally, we proposed a well-grounded strategy for training a surrogate loss to save computational complexity. We show that our model is significantly better than the baseline on two important tasks and that it is robust for several modeling choices.
	%Our results further share several interesting aspects, learning word embedding helps the models and scale of the data may influence the implications one can derive from experiments.
	
	% However, the proposed models were only allowed to use text data, while user, advertiser and other features were not used in this study. In the future work, we will explore ways of incorporating available additional features to evaluate the potential gain.
	
	%\section*{ACKNOWLEDGMENTS} This project was conducted while authors were interning/working at Yahoo Labs, Sunnyvale... or something like that?

	\balance
	\bibliographystyle{ACM-Reference-Format}
	\bibliography{sample-bibliography} 
	
\end{document}